\documentclass{article}
\usepackage[T1]{fontenc} 
\usepackage[utf8]{inputenc} 
\usepackage{ismir,amsmath,cite,url,amssymb}
\usepackage{microtype}
\usepackage{graphicx}
\usepackage{color}
\usepackage[table,xcdraw]{xcolor}

\usepackage{hhline}
\usepackage{multirow}

\title{vocadito: A dataset of solo vocals with $f_0$, note, and lyric Annotations}

\multauthor
{Rachel M. Bittner$^{\flat}$, Katherine Pasalo$^{\flat}$, Juan Jos\'e Bosch$^{\flat}$, Gabriel Meseguer-Brocal$^{\sharp}$, David Rubinstein$^{\flat}$}{ 
 $^{\flat}$Spotify, $^{\sharp}$IRCAM \\
}

\def\authorname{R. M. Bittner, K. Pasalo, J.J. Bosch, G. Meseguer Brocal, D. Rubinstein}

\begin{document}

\maketitle
\begin{abstract}
To compliment the existing set of datasets, we present a small dataset entitled \emph{vocadito}, consisting of 40 short excerpts of monophonic singing, sung in 7 different languages by singers with varying of levels of training, and recorded on a variety of devices.
We provide several types of annotations, including $f_0$, lyrics, and two different note annotations.
All annotations were created by musicians.
We provide an analysis of the differences between the two note annotations, and see that the agreement level is low, which has implications for evaluating vocal note estimation algorithms. 
We also analyze the relation between the $f_0$ and note annotations, and show that quantizing $f_0$ values in frequency does not provide a reasonable note estimate, reinforcing the difficulty of the note estimation task for singing voice.
Finally, we provide baseline results from recent algorithms on vocadito for note and $f_0$ transcription.
Vocadito is made freely available for public use.
\end{abstract}
\section{Introduction}\label{sec:introduction}
The singing voice is one of the most expressive instruments, and one that is particularly challenging to transcribe~\cite{humphrey2018introduction}.
A common task is to transcribe a voice's pitch content, either in the form of frame-level $f_0$, or in the form of note-events.
These two representations are related, but not trivial to convert between. 
Given a recording's frame-level $f_0$, one cannot trivially create note events by e.g. quantizing because it lacks information about onsets, and it is ambiguous how to group pitches into events.
Similarly, it is not possible to infer what the frame-level $f_0$ is for a recording given a sequence of notes, as expressive performance information such as vibrato or glissando are not encoded.

Notes themselves are known to have a degree of subjectivity - (as we'll see further results for in this paper) - given the same recording, two humans may not generate the same sequence of note events.
First, decision for where a note onset should be placed is also unclear - for a sung lyric, should an onset be placed at the start of an un-pitched transient, or at the start of the pitched vowel?
The center pitch of a note which contains a degree of fluctuation is yet another ambiguity, and there is a wide variance in what humans perceive as ``the'' pitch of a sung note~\cite{devaney2015influence,larrouy2018pitch,larrouy2018know}, and it depends often on context.
The notion of when a region should be split into two notes or should be a single note is often a matter of style or genre - some deviations can be considered as separate (e.g. grace) notes, or as simply part of a single note. 
Finally, the notion of a note offset is ambiguous, and even human annotations may show big discrepancies unless a very concrete annotation procedure is established ~\cite{liang2015musical}.

For this reason, when evaluating the correctness of estimated note events, it is common to allow a tolerance window for where an onset is placed, and an ever larger tolerance for where an offset is placed.
In other related tasks, e.g. chord recognition~\cite{de2011corpus} and music segmentation~\cite{smith2011salami}, datasets with multiple annotations have been created to address the inherent subjectivity of the task.
To date, no such dataset exists for note annotations.

Another interesting singing voice transcription task is transcribing lyrics at the word, syllable or even phoneme-level.
Additionally, there is a link between notes/$f_0$ and lyrics.
While this link (e.g. word-painting) is well studied in music theory~\cite{kroeger1988word,godt1984essay}, is under-explored computationally.

Few datasets exist which contain both human-annotated note and $f_0$ data, making it difficult to study interactions between them.
The same is true for lyrics and note or $f_0$ data.
Table~\ref{tab:datasets} gives an overview of existing datasets with solo or monophonic singing voice.
While there are a number of existing datasets with note, $f_0$ or lyric annotations, when it comes to any one task there are actually only a few.
When it comes to note estimation for note estimation for solo singing voice, only 3 exist: Molina, DALI\_multi, and TONAS\footnote{It is worth noting that ChoirSet is similar to vocadito in the annotations it provides (notes, $f_0$ and lyrics) - however it is primarily a polyphonic dataset.
While ChoirSet includes stems of individual singers, they contain bleed or artifacts due to the style of microphone, and are not well suited for monophonic voice evaluation.}.
DALI\_multi is large, however the annotations are crowdsourced and automatically aligned -- while this is useful for training, it is not an appropriate dataset for evaluation. Molina and TONAS are both good evaluations sets, but (like vocadito) are relatively small.
We are further restricted if both $f_0$ and note annotations are needed - leaving only TONAS.
Furthermore, no dataset provides more than one note annotation, and as we will see, the note annotation task itself is quite subjective.

In this report, we describe the creation of the \textit{vocadito} dataset.
We also provide an analysis of inter-annotator agreement for note annotations, and an analysis of how notes and quantized $f_0$ differ.
Finally, we report note estimation results by several algorithms on vocadito, and analyze open challenges.
Vocadito is made freely available on Zenodo\footnote{\url{https://zenodo.org/record/5557945}} under a Creative Commons license, and is included in the mirdata~\cite{mirdata_zenodo} library.





\begin{table*}
\centering
\begin{tabular}{r c c c c c c c} 
\hline
\textbf{Dataset} & \textbf{Polyphony} & \textbf{Isolated?}& \textbf{Notes?}  & \textbf{F0?} & \textbf{Lyrics?} & \textbf{Multi-annotation?} & \textbf{\# tracks}  \\
\hline
Saraga Carnatic  \cite{bozkurt_b_2018_4301737} & 1 &  &  & \checkmark &  & &249 \\
\textit{DALI}  \cite{meseguer2018dali,meseguer2020creating} & 1+ &  & \checkmark &  & \checkmark & &7756 \\
\textit{cante100}  \cite{nadine_kroher_2018_1322542, nadine_kroher_2018_1324183} & 1 &  & \checkmark & \checkmark &  & &100 \\
MIR-1K\footnote{\url{https://sites.google.com/site/unvoicedsoundseparation/mir-1k}} & 1 & \checkmark &  & \checkmark & \checkmark & & 1000 \\
\textit{iKala} \cite{chan2015vocal}  & 1 & \checkmark & & \checkmark & \checkmark & &252  \\
\textit{MedleyDB}  \cite{bittner2014medleydb,bittner2016medleydb} & 1 & \checkmark  &  & \checkmark & & & 93 \\
Jingju A Cappella  \cite{rong_gong_2018_1323561} & 1 & \checkmark &  &  & \checkmark & &82 \\
VocalSet \cite{wilkins2018vocalset} & 1 & \checkmark & & & \checkmark & & 3560\\
\textit{ChoirSet}  \cite{rosenzweig2020dagstuhl}  & 4+ & \checkmark & \checkmark & \checkmark  & \checkmark & &20 \\
\textit{DALI\_multi}  \cite{meseguer2020content} & 1+ & \checkmark & \checkmark &  & \checkmark & & 513 \\
\textit{Molina} \cite{molina2014evaluation} & 1 & \checkmark & \checkmark &   & & &38 \\
TONAS  \cite{tonas_music,tonas_annotations} & 1 & \checkmark & \checkmark & \checkmark &  & &72 \\

\hline
\hline 
\textit{vocadito}& 1 & \checkmark & \checkmark & \checkmark & \checkmark & \checkmark & 40  \\
\hline
\end{tabular}
\caption{A (non-exhaustive) overview of existing datasets for solo vocals. \textbf{Polyphony} indicates how many voices are present at one time. A $+$ indicates that there are recordings where multiple singers are singing at once. \textbf{Isolated} indicates whether the vocal recordings are isolated or not (there is background music).}
\label{tab:datasets}
\end{table*}

\section{Dataset Creation}

\subsection{Data Collection}

Audio recordings for \textit{vocadito} were collected from 28 volunteers, with varying singing experience.
In order to simulate a ``real-world'' setting, we did not restrict volunteers to record using high-quality microphones, and many of the recordings are from cell phone or computer microphones.
Volunteers were asked to choose an original or public domain song (e.g. folk or children's music), and create up to three 10-40 s recordings.
Volunteers agreed to their recordings being anonymously included in this dataset and publicly released.
We ensured that no composition is repeated across the 40 recordings.

The collected recordings were manually edited to remove any long silences at the beginnings or ends, and reformatted from their original format into 44.1 kHz, 16 bit mono .wav files.

\subsection{Human Annotations} 
We created four types of human-labeled annotations for Vocadito: frame-level $f_0$, notes, lyrics, and track-level metadata (e.g. the sung language).
All annotators are experienced musicians.

\subsubsection{$f_0$}
$f_0$ annotations are created using Tony~\cite{mauch_computer-aided_nodate}, a software application which first automatically estimates the $f_0$ using the pYIN~\cite{mauch2014pyin} algorithm, and then allows an annotator to manually correct mistakes made by the algorithm.
One annotator created $f_0$ annotations in this manner for each of the 40 tracks.
The annotator reported that the majority of the corrections involved either removing $f_0$ estimates in frames where no $f_0$ was present (e.g. during consonants), or in adding missing $f_0$ estimates for frames with low pitch.

\subsubsection{Notes}
Note annotations were also created using Tony, which similarly for $f_0$, uses an algorithm to estimate note events and allows an annotator to correct the estimates.
In order to explore the subjectivity of creating note events for vocals, two different annotators created separate note event annotations for each of the 40 tracks.
The annotators were instructed to annotate the notes they would play if they had to reproduce it on the piano.
Note that while the piano is restricted to a semitone grid, the annotation software (Tony) allows notes to have any continuous pitch; which is convenient, since singers did not necessarily sing in standard tuning (440 Hz).
Thus, the instructions provided to the annotators regarding the piano refers more to how to segment notes in time than to how to label the pitch.

\subsubsection{Lyrics}
Lyric annotations were created by fluent speakers of the sung language for each song.
Annotators listened to the recording and transcribed the words as they are sung exactly (even when this deviates slightly from the text of the original composition).
Line breaks in the lyrics indicate the end of a musical phrase, and a blank line indicates the end of a musical section.
The lyrics are provided as text, without timing information.
All lyrics are written in the Latin alphabet. 
For the two tracks which are in Chinese, lyrics are provided in both Chinese characters and in pinyin.
For the two tracks where more than one language is present, we indicate the language as \texttt{language1+language2}.

\subsubsection{Metadata}
We provide track-level metadata for each of the tracks, including (1) the sung language (2) the singer ID (anonymized) (3) the average pitch (computed from the $f_0$ annotations).

\subsection{Dataset Statistics}

\begin{figure}[h]
    \centering
    \includegraphics[width=\columnwidth]{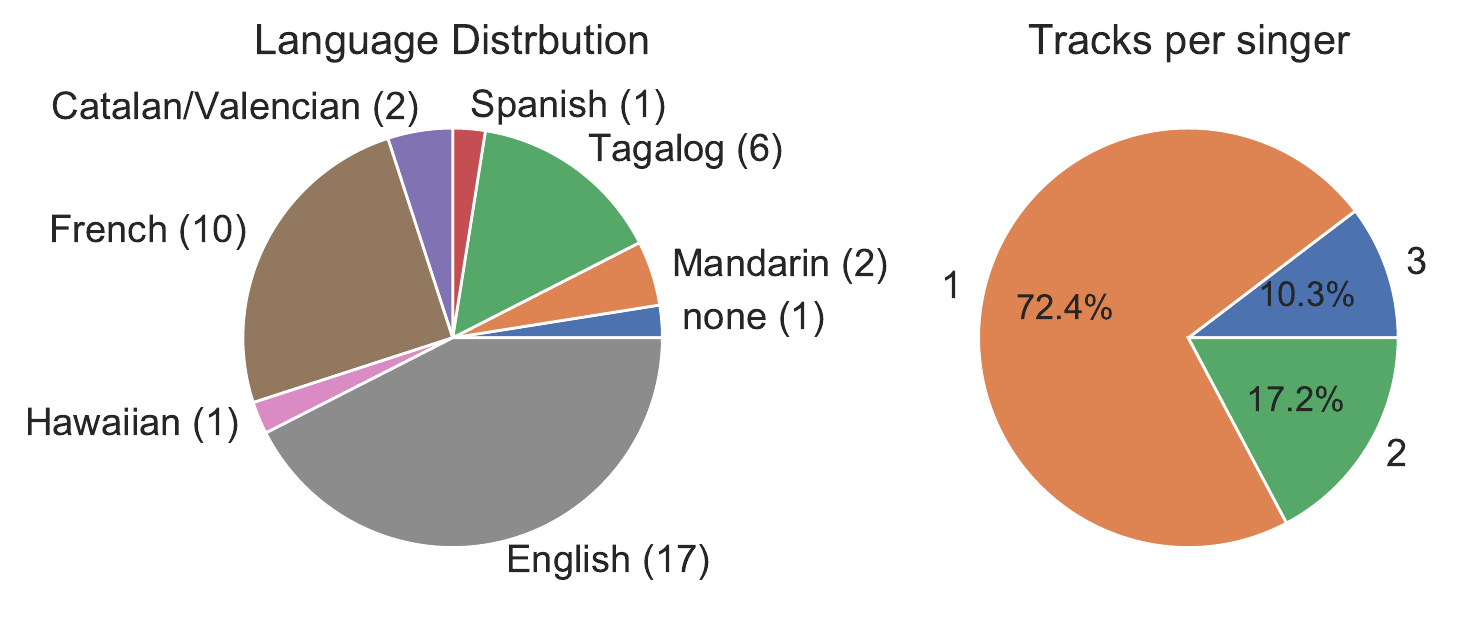}
    \caption{(Left) The distribution of languages in vocadito. Track counts are shown in parentheses. (Right) The distribution of tracks per singer.}
    \label{fig:track_dist}
\end{figure}

Vocadito has a total of 40 tracks, consisting of 29 unique singers, singing in 7 different languages.
The distribution of languages and number of tracks per singer is illustrated in Figure~\ref{fig:track_dist}.
English and French dominate the language distribution, but there are tracks in 5 additional languages present as well.
Note there is one track with no labeled language, as the singer simply sings ``la'' throughout the excerpt.
The majority of singers are in only one track, however there are a few singers who submitted 2 or 3 tracks.

\begin{figure}[h!]
    \centering
    \includegraphics[width=\columnwidth]{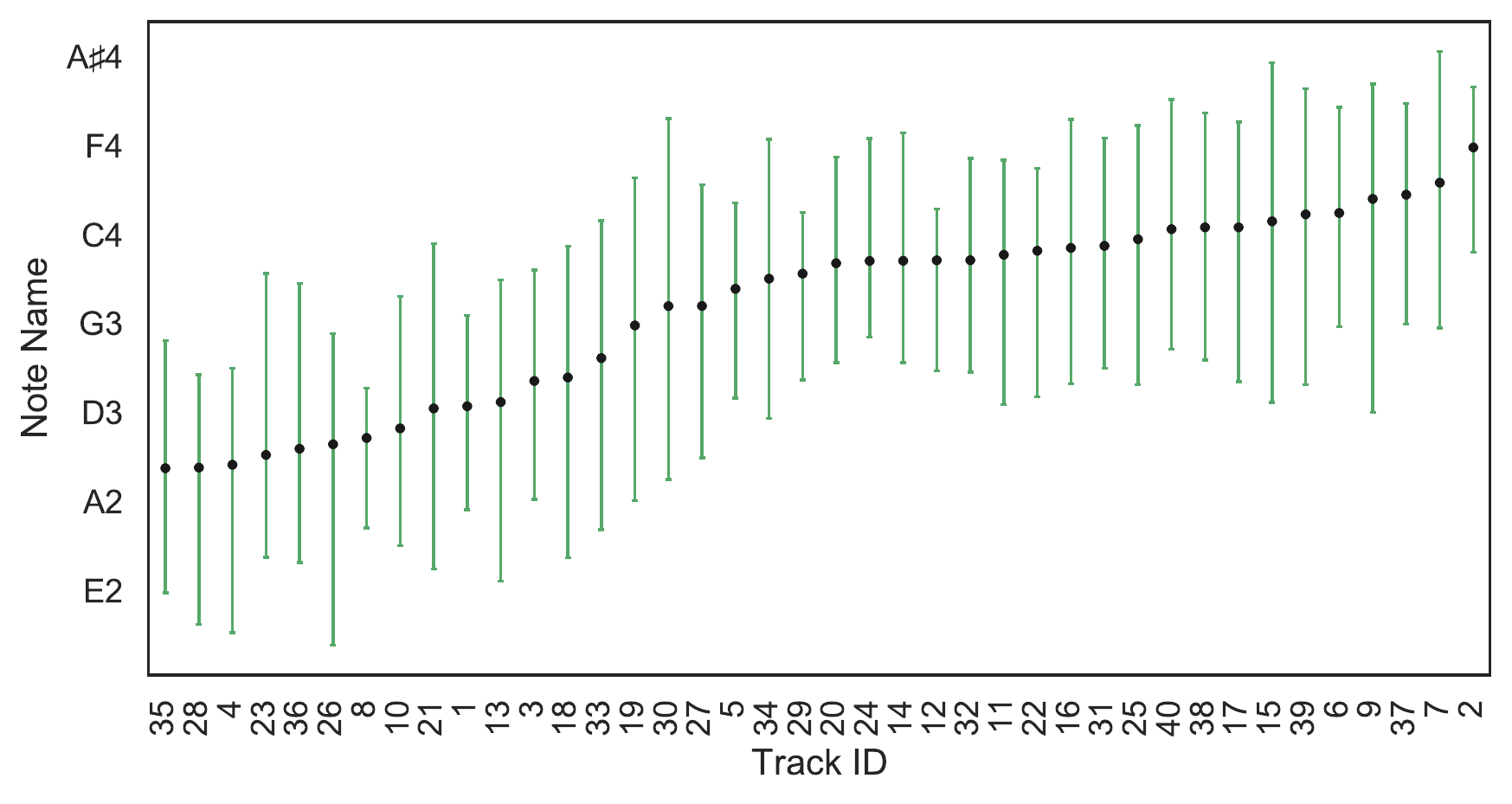}
    \caption{Mean, minimum and maximum $f_0$ values (as note names) for \textit{vocadito}, ordered by average $f_0$. The x-axis indicates the corresponding vocadito track ID.}
    \label{fig:pitch_ranges}
\end{figure}

In terms of pitch register, we analyze each singer's average pitch (computed from the $f_0$ annotations, in midi-note units).
Figure~\ref{fig:pitch_ranges} shows the pitch range of each song, ordered by the average pitch from lowest to highest.
The tracks cover the bass, tenor and alto ranges well, but there are few tracks that are in the soprano singing range.

\section{Experiments}

\subsection{Subjectivity of Vocal Notes}\label{sec:annotation-diff}

Following previous related works \cite{bosch2014melody, rosenzweig2021reliability}, we analyze inter-annotator agreement by computing several metrics using pairs of annotations as either estimation or reference.
We report the frame-level metric \texttt{Accuracy} (\texttt{Acc}), which first converts the note labels into 10 ms frame-wise $f_0$ values, and computes the percentage of time frames with matching labels.
For pairs of frames which both have a pitch estimated, a frame is considered matching if the two $f_0$ values are within a semitone of one-another.
We also report a set of standard note event-level transcription metrics~\cite{mcleod2018evaluating}, including \texttt{F-score} (\texttt{F}) and the equivalent metric where offsets are not considered (\texttt{F}$_{no}$) computed over note events.
In practice, since we only have two annotators, we just need to compute metrics using one of the annotators as reference. When using the second annotator as reference the resulting metrics are equivalent with Precision/Recall inverted, making the two F-scores identical in both cases.
Similarly, Accuracy is also equivalent in both, as long as the number of reported frames is equivalent.

The agreements (using Annotator 2 as the reference) are shown in Table~\ref{tab:annotator_agreement}.
The low average value for \texttt{F} (0.64) demonstrates how truly subjective this task is.
Additionally, the track for which the annotators agreed the least has \texttt{F}=0.34, and only 0.88 the track for which they agreed the most.
If we now ignore differences in offset and look at \texttt{F}$_{no}$, there is still only 0.74 agreement.
When considering \texttt{Acc}, the agreement is slightly higher (0.80 on average), however this is likely because the differences in onsets/offsets are not penalized as heavily as in \texttt{F}.

\begin{table}[h]
    \centering
    \begin{tabular}{|r c c c c|}
        \hline
       \textbf{Metric}  & $\mu$ & $\sigma$ & min & max \\
       \hline
        \texttt{Acc} & 0.83 & 0.08 & 0.61 & 0.92\\
        \texttt{F} & 0.64 & 0.13 & 0.34 & 0.88\\
        \texttt{F}$_{no}$ & 0.74 & 0.09 & 0.57 & 0.94\\
        \hline
    \end{tabular}
    \caption{Agreement between note annotations, using Annotator 2 as the reference. Note that using Annotator 1 as the reference would result in the same scores $F$ and $F_no$ scores. We report the mean ($\mu$), standard deviation ($\sigma$), minimum (min) and maximum (max) over the 40 tracks.}
    \label{tab:annotator_agreement}
\end{table}

\begin{figure}[h]
    \centering
    \includegraphics[width=\columnwidth]{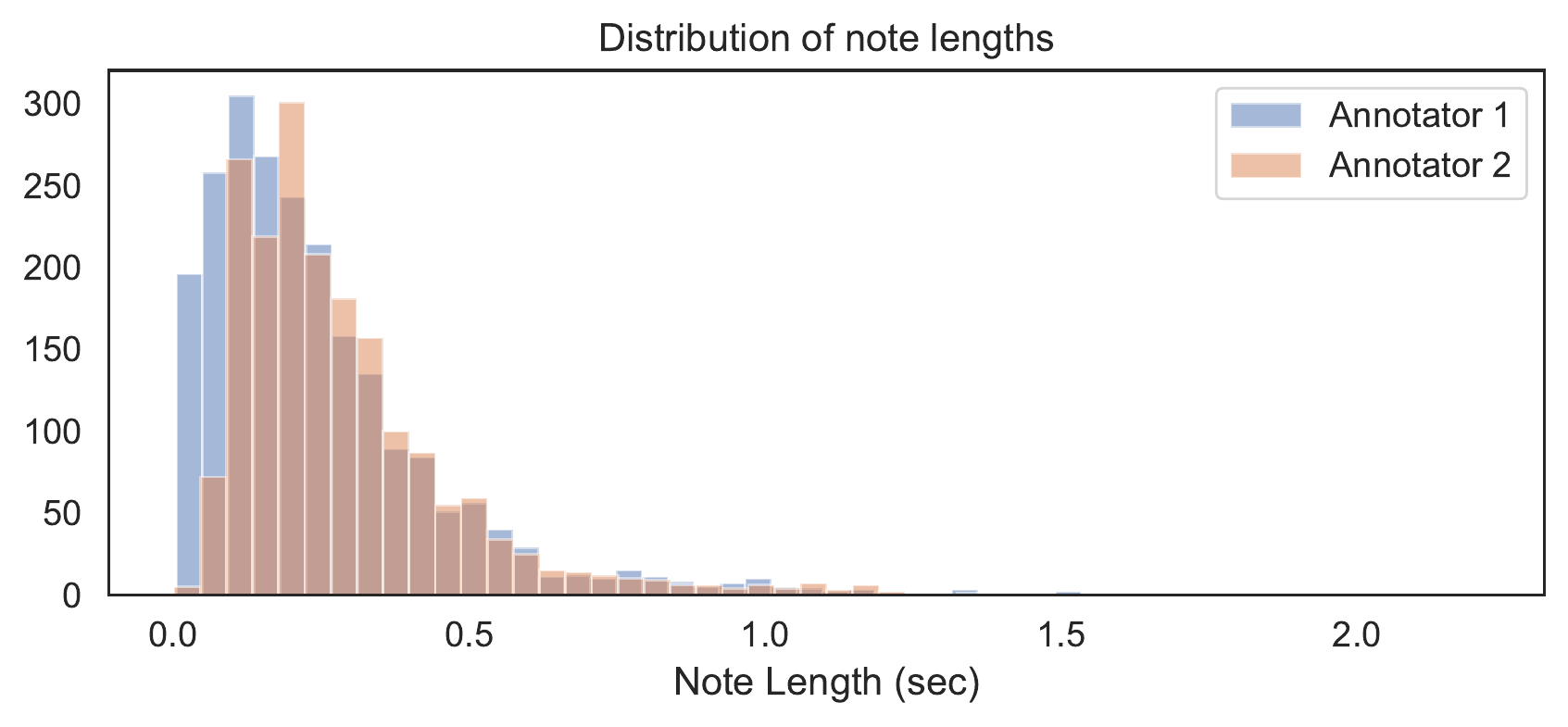}
    \caption{Distribution of note lengths for Annotators 1 and 2. Annotator 1 annotated smaller deviations in pitch (e.g. grace notes, ornaments) as separate notes, while Annotator 2 grouped ornaments into single notes. This is reflected by the large difference in the number of short notes.}
    \label{fig:note_lengths}
\end{figure}

To better understand the nature of the differences in annotation, we first consider the distribution of note lengths for Annotators 1 and 2, plotted in Figure~\ref{fig:note_lengths}.
We see that the largest discrepancy is in the number of short notes, with Annotator 1 labeling a large number of short notes, and Annotator 2 labeling very few.
Upon further examination, we see that Annotator 1 labeled more grace notes/ornaments as individual notes, while Annotator 2 grouped them into single longer notes.

\begin{figure*}
    \centering
    \includegraphics[width=2\columnwidth]{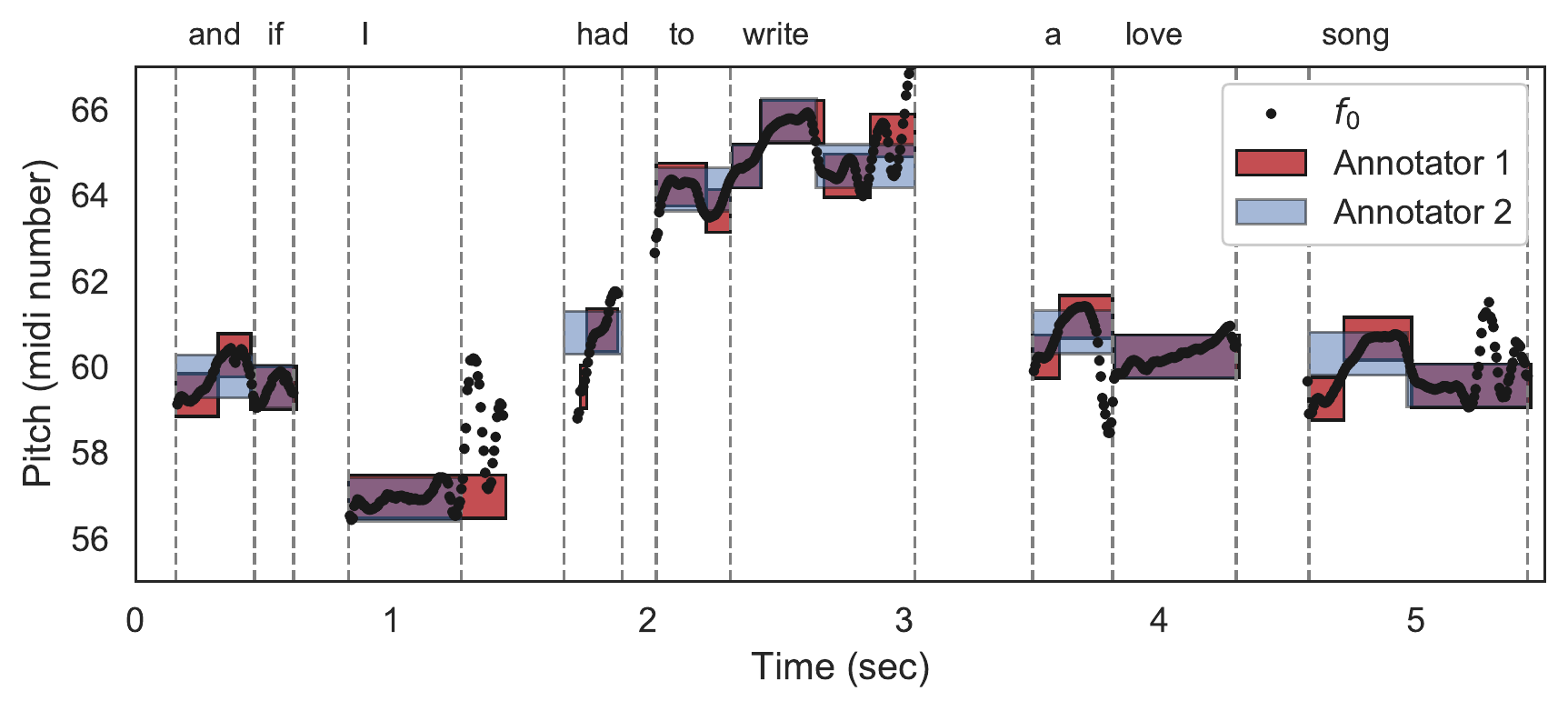}
    \caption{Annotations for the first 5.5 seconds of track 6 of vocadito. The plot shows $f_0$ annotations (black), note annotations by annotators 1 and 2 in red/blue respectively (overlaps shown in purple), and lyrics above the plot. Lyric time-alignments were labeled here for demonstration purposes, but are not part of vocadito.}
    \label{fig:annotator_differences}
\end{figure*}

We further investigate this difference with a qualitative example, shown in Figure~\ref{fig:annotator_differences}.
Here we see the notes for each annotator, alongside the lyrics and annotated $f_0$ for the first 5 seconds of a track.
Here we see a confirmation of our observation: Annotator 1 labeled grace notes and vocal ornaments as separate notes, while Annotator 2 grouped them into single notes.
For example, the first lyric ``and'' has two notes annotated by Annotator 1 and one by Annotator 2.
We note that Annotator 1 segmented notes based more on the degree of pitch fluctuation, while Annotator 2 segmented notes based more on the lyrical content.
Another discrepancy we see is in how note start and end times are interpreted.
The notes corresponding to the lyrics ``I'' have a large discrepancy in their offset, where Annotator 1 interpreted the ending, quiet ornament as part of the note and Annotator 2 did not.
The notes corresponding to the lyric ``had''  have a discrepancy in the timing of the onsets, where Annotator 2 notated the onset along with the lyric start (annotating the transient as part of the note), while Annotator 2 annotates it as starting when the pitch begins. 

We remark that we do not consider either annotator to be right or wrong, and merely outline the differences in interpretation to highlight how subjective this task is.

\subsection{Notes vs quantized f0}

We perform a simple experiment to demonstrate the difference between human-annotated note events and note events which are trivially generated from the $f_0$ annotation.
We create ``automatic'' note events by quantizing every time frame in the $f_0$ annotation to the nearest semitone.
All consecutive sequences of the same pitch define a note, where the start time is the time stamp of the first frame of the sequence, and the end time is the time stamp of the frame following the sequence.
To evaluate how this compares to the human annotations, we compute \texttt{Acc}, \texttt{F}, and \texttt{F}$_{no}$ using both human annotations as a reference and taking the score providing the highest value.
The resulting metrics have an average \texttt{Acc} of 0.55 ($\sigma$=0.09), \texttt{F} of 0.20 ($\sigma$=0.07), and \texttt{F}$_{no}$ of 0.38 ($\sigma$=0.06).
We see that the quality of these ``automatic'' note event annotations (derived from ground truth $f_0$) are not at all accurate, reinforcing the non-triviality of generating note event annotations.

\subsection{Baseline Results}

\subsubsection{Notes}
Here we provide baseline results for a note estimation algorithm \textit{Vocano}~\cite{hsu_vocano} on vocadito.
Vocano uses a pre-trained pitch tracking model and applies a neural network to segment the $f_0$ into note events which is trained on the TONAS~\cite{Mora_2010} and MIR-1K\cite{hsu2009improvement} datasets.

We compute results per-annotator (A$_1$, A$_2$) and combined (A$_{max}$), shown in Table~\ref{tab:note_baseline}.
Since both annotations are valid, we combine the two annotations by choosing, for each track, the annotation that maximizes \texttt{F} and then computing the other metrics based on the selected annotation, followed by computing a dataset-level average.

\begin{table}[h]
    \centering
    \begin{tabular}{|r c c c|}
        \hline
        \textbf{Metric} & \textbf{A}$_1$ & \textbf{A}$_2$ & \textbf{A}$_{max}$ \\
        \hline
        \texttt{Acc} & 0.55 (0.11) & 0.55 (0.10) & 0.56 (0.10) \\
        \texttt{F} & 0.43 (0.11) & 0.49 (0.10) & 0.50 (0.10) \\
        \texttt{F}$_{no}$ & 0.57 (0.11) & 0.63 (0.09) & 0.64 (0.09) \\
        \hline
    \end{tabular}
    \caption{Note Metrics for the Vocano algorithm. The A$_1$ and A$_2$ columns show the mean (and standard deviation in parentheses) metrics compared with Annotators 1 and 2 respectively. The A$_{max}$ column shows the average over tracks when using the metrics compared with the annotator giving the highest \texttt{F}.}
    \label{tab:note_baseline}
\end{table}

For this algorithm, we see that the scores are slightly higher for Annotator 2's transcriptions, and (by nature) the A$_{max}$ scores are higher than for either annotator alone.
While the scores appear low in absolute, we see that, compared to the level of annotator disagreement, the \texttt{F} score is only 14 percentage points below the annotator disagreement level (0.64).
When reporting results on vocadito, if any style of transcription is acceptable, we recommend reporting the scores as in A$_{max}$. 
If instead the application requires either fine-grained or a coarse transcription, scores could be reported for Annotators 1 or 2 only, respectively.

\subsubsection{$f_0$}

\begin{table}[h!]
    \centering
    \begin{tabular}{|r c c|}
        \hline
        \textbf{Metric} & $\mu$ & $\sigma$ \\
        \hline
        Voicing Recall & 0.89 & 0.03\\
        Voicing False Alarm & 0.31 & 0.08\\
        Raw Pitch Accuracy & 0.98 & 0.02\\
        Raw Chroma Accuracy & 0.98 & 0.01\\
        Overall Accuracy & 0.82 & 0.04\\
        \hline
    \end{tabular}
    \caption{Single-$f_0$ metrics for \textit{crepe} on vocadito.}
    \label{tab:f0_baseline}
\end{table}

Finally, we report results for the popular \textit{crepe}~\cite{kim_crepe_2018} pitch tracking algorithm on vocadito.
We report the metrics described in~\cite{bittner2019generalized}, using the confidence output from crepe as a continuous voicing output.
The results in Table~\ref{tab:f0_baseline} show that while the accuracy is quite high, there is some remaining margin when it comes to estimating when a pitch is or is not active.
Impressively, the pitch accuracy is nearly perfect, and the algorithm makes virtually no octave mistakes.

\section{Conclusions}
In this report, we introduced \textit{vocadito}, a small dataset of 40 recordings of solo voice.
We described the annotation procedure and provided dataset-level statistics.
Through our analysis of multiple note annotations, we saw the degree to which annotating vocal notes can be highly subjective, with only 0.64 F-measure agreement on average.
Furthermore, we showed that simply quantizing ground-truth $f_0$ values provides note estimates which are far from either human annotator's transcription, emphasizing the need for dedicated algorithms for the task of vocal note transcription.
Finally, we provided baseline results for $f_0$ and note transcription.

Future work could explore the interactions between vocal note, $f_0$ and lyric annotations, in order to better understand the computational links between them.
For example, it is not well understood (computationally) exactly where a vocal note onset or offset should occur and why. Several heuristics have been investigated~\cite{mauch_computer-aided_nodate, panteli2017towards, kroher2016automatic}, such as modeling large fluctuations in energy/spectral flux, lyrical transients, or in pitch. However, there has historically been a lack of data in this domain, and so the existing experimental results are necessarily limited.

The choice of the ``center'' pitch value of a note given a sequence of $f_0$ curves is also still an open question.
Heuristics also exist for this task, such as choosing the median pitch value, or modeling (and removing) vibrato and glissando before selecting the pitch value \cite{kroher2016automatic, molina2014sipth,panteli2017towards}.
The links between the computational work in MIR and in music psychology on this topic can be strengthened, as there may remain a degree of subjectivity in this task as well~\cite{devaney2015influence,larrouy2018pitch,larrouy2018know}. 

A missing link in the annotations provided in vocadito are lyric alignments -- while lyrics are provided, we do not align them to the audio files. This annotation extension would add another analysis dimension possibility to vocadito.

Finally, note transcription metrics are not designed for more than one annotation. While we have provided a preliminary recommendation (for each track using the annotation which gives the highest F-score), the metrics themselves could be expanded to better handle ``multiple correct answers'' over time.

\bibliography{references}

\end{document}